# Low-loss photonic crystal fibers for transmission systems and their dispersion properties

M.D. Nielsen[1,2], C. Jacobsen[1], N.A. Mortensen[1], J.R. Folkenberg[1], and H.R. Simonsen[1]

[1] *Crystal Fibre A/S, Blokken 84, DK-3460 Birkerød, Denmark*
[2] *COM, Technical University of Denmark, DK-2800 Kongens Lyngby, Denmark*

*mdn@crystal-fibre.com*

**Abstract:** We report on a single-mode photonic crystal fiber with attenuation and effective area at 1550 nm of 0.48 dB/km and 130 $\mu m^2$, respectively. This is, to our knowledge, the lowest loss reported for a PCF not made from VAD prepared silica and at the same time the largest effective area for a low-loss (< 1 dB/km) PCF. We briefly discuss the future applications of PCFs for data transmission and show for the first time, both numerically and experimentally, how the group velocity dispersion is related to the mode field diameter.
©2004 Optical Society of America

## References and Links

1. K. Tajima, K. Nakajima, K. Kurokawa, N. Yoshizawa, and M. Ohashi "*Low-loss photonic crystal fibers,*" Optical fiber communications conference, OFC 2002 (Anaheim, CA, USA)**,** pp. 523-524 (2002).
2. L. Farr, J. C. Knight, B. J. Mangan, and P. J. Roberts "*Low loss photonic crystal fibre,*" 28th European conference on optical communication (Copenhagen, Denmark)**,** PD1-3, (2002).
3. K. Tajima, J. Zhou, K. Nakajima, and K. Sato "*Ultra low loss and long length photonic crystal fiber,*" Optical fiber communications conference, OFC (Anaheim, CA, USA)**,** PD1, (2003).
4. K. Tajima, J. Zhou, K. Kurokawa, and K. Nakajima "*Low water peak photonic crystal fibers,*" 29th European conference on optical communication ECOC'03 (Rimini, Italy)**,** pp. 42-43 (2003).
5. C.M. Smith, N. Venkataraman, M.T. Gallagher, D. Müller, J.A. West, N.F. Borrelli, D.C. Allan, and K.W. Koch, "Low-loss hollw-core silica/air photonic bandgap fibre," Nature **424**, 657-659 (2003).
6. N.A. Mortensen and J.R. Folkenberg, "Low-loss criterion and effective area considerations for photonic crystal fibers," J. Opt. A: Pure Appl. Opt. **5**, 163-167 (2003).
7. O. Humbach, H. Fabian, U. Grzesik, U. Haken, and W. Heitmann, "Analysis of OH absorption bands in synthetic silica," J. Non-Cryst. Solids **203**, 19-26 (1996).
8. M.D. Nielsen and N.A. Mortensen, "Photonic crystal fiber design based on the V-parameter," Opt. Express **11**, 2762-2768 (2003). http://www.opticsexpress.org/abstract.cfm?URI=OPEX-11-21-2762
9. M.D. Nielsen, J.R. Folkenberg, N.A. Mortensen, and A. Bjarklev, "Bandwidth comparison of photonic crystal fibers and conventional single-mode fibers," Opt. Express **12**, 430-435 (2004). http://www.opticsexpress.org/abstract.cfm?URI=OPEX-12-3-430
10. M.D. Nielsen, J.R. Folkenberg, and N.A. Mortensen, "Singlemode photonic crystal fiber with effevtive area of 600 $\mu m^2$ and low bending loss," Electron. Lett. **39**, 1802-1803 (2004).
11. S.G. Johnson and J.D. Joannopoulos, "Block-iterative frequency-domain methods for Maxwell's equations in a planewave basis," Opt. Express **8**, 173-190 (2001). http://www.opticsexpress.org/abstract.cfm?URI=OPEX-8-3-173
12. J. Lægsgaard, A. Bjarklev, and S.E.B. Libori, "Chromatic dispersion in photonic crystal fibers: fast and accurate scheme for calculation," J. Opt. Soc. Am. B **20**, 443-448 (2003).
13. G.P. Agrawal, *Fiber-Optic Communication Systems* (John Wiley & Sons, Inc., 1997)
14. T. Kato, M. Hirano, M. Onishi, and M. Nishimura, "Ultra-low nonlinearity low-loss pure silica core fiber for long-haul WDM transmission," Electron. Lett. **35**, 1615-1617 (1999).
15. K. P. Hansen, J. R. Jensen, C. Jacobsen, H. R. Simonsen, J. Broeng, P. M. W. Skovgaard, A. Petersson, and A. Bjarklev "*Highly Nonlinear Photonic Crystal Fiber with Zero-Dispersion at 1.55 $\mu m$,*" Optical fiber Communications conference OFC 2002 (Anaheim, CA, USA)**,** (2002).

## 1. Introduction

In recent years, the typical attenuation level of photonic crystal fibers (PCFs) has been reduced dramatically. This is both true for fibers relying on index guiding [1-4] as well as those based on the photonic bandgab effect [5], although the latter type still needs to improve with almost two orders of magnitude in order to have lower loss than the index guiding fibers.

A few years back, the typical attenuation level for index guiding PCFs was a few dB/km at 1550 nm. In the beginning of 2002 the 1 dB/km limit was reached [1] and during a relatively short period of time, this was improved to the current record level of 0.28 dB/km [4]. The most recent improvements have been obtained by using high purity glass prepared by the vapour-phase axial deposition (VAD) technique and by eliminating the presence of OH contamination. While these advances are indeed both significant and impressing, most of the fibers reported have relative small effective areas no larger than ∼80 µm$^2$ [4], and typically ∼20 µm$^2$ [1-3]. Since the sensitivity towards attenuation caused by structural variations, both in the transverse and longitudinal dimension of the fiber, increases as the effective area is increased, it is interesting to hold up the effective area against the obtained attenuation level [6].

Here we present results on a low-loss PCF with an effective area significantly larger than that of previous reports. Also, we numerically investigate the relation between the mode-field diameter (MFD) and the group velocity dispersion (GVD) at the 1550 nm wavelength and compare the obtained results with measurements performed on a broad range of PCFs with widely different dimensions. All the fibers considered here are index-guiding PCFs with a cladding region consisting of a triangular arrangement of air holes running along the full length of the fiber surrounding a central core region which is formed by omitting a single air hole.

## 2. Fabricated fiber

An optical micrograph of the fabricated PCF is shown in Fig. 1, indicating the fiber diameter, $D = 173$ µm, the air-hole diameter, $d = 5.15$ µm, and the pitch, $\Lambda = 10.6$ µm. The fiber has a single-layer acrylate coating (not shown) with a diameter of 315 µm. As seen from the picture, the microstructured cladding region consists of 54 air holes corresponding to 4 periods of which the 6 holes in the corners of the outer ring have been omitted.

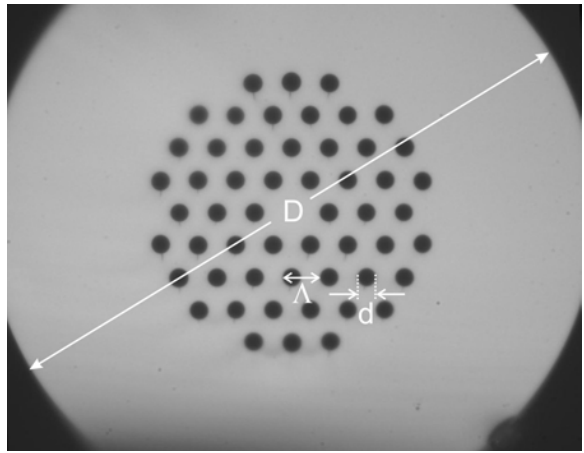

Fig. 1. Optical micrograph of the fabricated PCF.

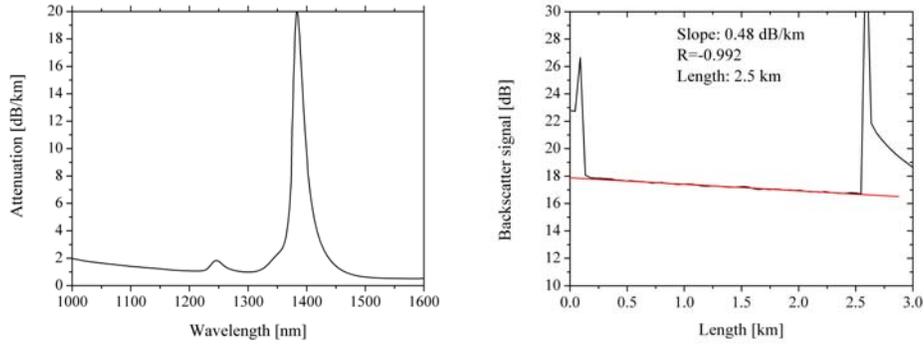

Fig. 2. Measured spectral attenuation (left) and OTDR trace at 1550 nm (right) of the fabricated PCF.

The fiber was fabricated by the stack and pull method where capillary tubes made of pure silica are stacked in a hexagonal pattern around a central silica rod. The surfaces of the elements were polished prior to stacking in order to reduce the influence of surface roughness [1]. The raw materials used for this fiber were commercially available silica rods and tubes with a typical OH content in the order of ∼0.5 ppm.

The spectral attenuation of the fiber was measured by the cutback technique using a white light source and an optical spectrum analyzer. The fiber length used for this measurement was 2.5 km and the obtained attenuation spectrum is shown in Fig. 2 (left). In order to check for inhomogeneities, such as scattering points, along the fiber length, the fiber was inspected using an optical time domain reflectometer (OTDR) with an operating wavelength of 1550 nm and pulse duration of 10 ns. The obtained OTDR trace is shown in Fig. 2 (right) as the black solid line. The red line, also shown in Fig. 2 (right), is the result of a linear regression on the measured data yielding a slope of -0.48 dB/km with a Pearson factor, $R = -0.992$, indicating a highly homogeneous fiber. The agreement between the attenuation at 1550 nm obtained from the spectral measurements and from the OTDR is excellent and well within the measurement uncertainty.

Analyzing the spectral attenuation data shows an OH induced attenuation peak at 1380 nm in the order of 20 dB/km corresponding to an OH concentration of ∼0.4 ppm [7] and an attenuation contribution at 1550 nm of 0.15 dB/km [4]. This is in good agreement with the expected values from the raw materials. By extracting the $\lambda^{-4}$ dependent scattering component, the Rayleigh scattering coefficient was determined to be 1 dB/(km·µm$^4$), equivalent to a contribution of 0.18 dB/km at 1550 nm. The remaining 0.16 dB/km of the attenuation at 1550 nm is attributed to the absorption from impurities and other imperfections.

From $d/\Lambda$ and $\Lambda$, the effective area of the fundamental mode is calculated to be 130 µm$^2$ [8] which is equivalent to a Gaussian MFD of ∼13 µm. This way of determining the MFD has proven to agree with measured data well within the typical measurement uncertainty [9,10]. This is to our knowledge the largest effective area reported so far for a low loss (< 1dB km) PCF.

## 3. Dispersion

As the effective area is increased and the strength of the wave guiding decreases, the GVD is expected to approach that of the bulk material [10]. Since the waveguide contribution to the dispersion is always positive when $\lambda \ll \Lambda$, the GVD must decrease towards the material dispersion for increasing effective area.

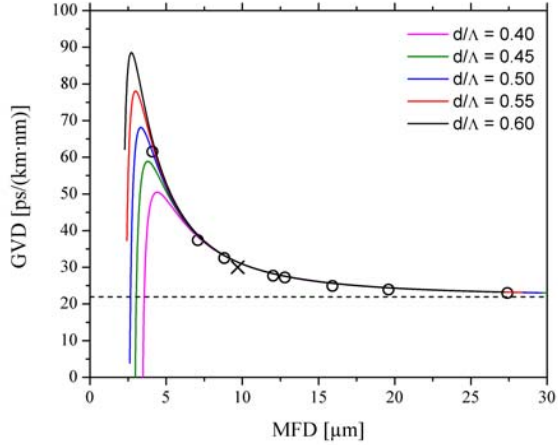

Fig. 3. GVD at 1550 nm as function of the MFD. The solid line indicates numerical results and the dashed line indicates $D_M$. Measured values at 1550 nm for 8 different PCF are indicated by circles while the cross represents independent data from Ref. [4].

By using the plane-wave expansion method [11], the effective index, $n$, as function of the free-space optical wavelength, $\lambda$, and the effective area of the fundamental mode was calculated for a range of $d/\Lambda$ values from 0.40 to 0.60 covering the relevant parameter space for possible transmission fibers. In these numerical simulations, the refractive index of silica was taken to have a constant value of 1.444. The GVD is defined as the derivative of $v_g^{-1}$ with respect to $\lambda$, where $v_g$ is the group velocity, and can therefore be calculated by the following relation where $c$ is the speed of light and $D_M$ is the material dispersion:

$$GVD = \frac{d}{d\lambda}\left(\frac{1}{v_g}\right) = -\frac{\lambda}{c}\frac{d^2 n}{d\lambda^2} + D_M$$

In this expression it has been assumed that the waveguide contribution to the GVD is independent of $D_M$ which is a good approximation provided that $\lambda \ll \Lambda$ [12]. The value of $D_M$ for pure silica at 1550 nm was calculated from the Sellmeier relation [13] to be 21.9 ps/(km·nm). In Fig. 3, the calculated GVD at 1550 nm is plotted as function of the MFD for $d/\Lambda$ = 0.40, 0.45, 0.50, 0.55, and 0.60 indicated by solid lines while $D_M$ is indicated by the dashed horizontal line. As seen from this plot, all the solid curves coincide for MFDs larger than ∼5 µm and the GVD is thus in this range given by the MFD regardless of the exact value of $d/\Lambda$. To verify whether the calculated relation corresponds with measurements, the group delay as function of wavelength was measured for a broad range of PCFs with widely different MFDs and from these data the GVD at 1550 nm of each fiber was derived. The measured fibers all had the same basic structure although the number of air holes varied from fiber to fiber. The MFDs of all the characterized fibers were calculated from the relation given in Ref. [8] and in Table 1, $\Lambda$, $d/\Lambda$, calculated MFD and measured GVD are listed. The number in the IDs refers to the approximate size of the core region. The measurements listed in Table 1 are indicated in Fig. 3 by open circles and in general, excellent agreement with numerical results is observed with a typical deviation of 0.5 ps/(km·nm). The deviation for the LMA5 is seen to be slightly larger which is attributed to the exact determination of $d/\Lambda$ which plays a role for this fiber only and, more importantly, to the fact that the assumption that $\lambda \ll \Lambda$ is no longer fulfilled. The data point in Fig. 3 indicated by a cross represents independent data from Ref. [4] and also in this case the agreement is very good.

Table 1. Structural parameters, calculated MFD and measured dispersion of the tested PCFs

| ID | $\Lambda$ [μm] | $d/\Lambda$ | MFD [μm] | GVD [ps/(km·nm)] |
|---|---|---|---|---|
| LMA-5[1] | 2.9 | 0.44 | 4.1 | 61.5 |
| LMA-8[1] | 5.6 | 0.46 | 7.1 | 37.4 |
| LMA-10 | 7.2 | 0.48 | 8.8 | 32.5 |
| LMA-15 | 10.0 | 0.50 | 12.0 | 27.7 |
| LMA-16 | 10.6 | 0.49 | 12.8 | 27.2 |
| LMA-20 | 13.0 | 0.47 | 15.9 | 24.9 |
| LMA-25 | 16.4 | 0.50 | 19.5 | 23.9 |
| LMA-35[2] | 23.2 | 0.50 | 27.4 | 23.0 |

[1] More information on this fiber is available in Ref. [9]
[2] More information on this fiber is available in Ref. [10]

## 4. Discussion and conclusion

The recent advances in the reduction of the attenuation level of index-guiding PCFs have left little doubt that these fibers, from an attenuation point of view, will be able to compete with conventional solid fibers for data transmission applications. One can even speculate that the ultimate attenuation level of the PCF might be even lower than that of conventional fibers since the PCF is a single material fiber with no boundary between two types of glass with different thermal expansion coefficients. However, little interest has until now been paid towards the GVD properties of these fibers in the telecom window. As demonstrated here, the GVD for the PCF will always tend to have a higher value than what is typical for a conventional solid fiber with comparable MFD. With proper dispersion compensation, this high GVD could be an advantage since it tends to suppress nonlinear interaction between channels in multi-wavelength transmission systems [14].

The numerical results in Fig. 3 have a slight offset compared with the measured data in the order of 0.5 ps/(km·nm). Since this offset is independent of both the MFD and the GVD values, we attribute this to a corresponding deviation in calculated material dispersion from the actual material dispersion value of the glass.

For sufficiently small MFDs, the GVD passes through zero and which has been used for realization of nonlinear fibers with zero dispersion at 1550 nm [15]. At some point, a further reduction of the structural scale, the mode will begin to expand relative to the structure and the MFD will increase. The simple behavior outlined here for large-mode area fibers should therefore not be extrapolated to nonlinear fibers.

We have demonstrated a PCF with an effective area of 130μm$^2$ and attenuation at 1550 nm of 0.48 dB/km and thereby shown that PCFs with effective areas able to compete with conventional solid fibers can be obtained while keeping the attenuation low. A significant contribution to this attenuation level was OH contamination present in the raw materials. We believe that the attenuation can be reduced close to the fundamental limit of pure silica while, at the same time, even larger effective areas than the one reported here, can be obtained.

## Acknowledgements

M.D. Nielsen acknowledges financial support from the Danish Academy of Technical Sciences.